\documentclass[nopacs,twocolumn,aps,pre,superscriptaddress]{revtex4-1}

\usepackage{textcomp}
\usepackage{amsmath}
\usepackage{amssymb}
\usepackage{mathrsfs}
\usepackage[normalem]{ulem}
\usepackage{color}
\usepackage{empheq}
\usepackage{multirow}
\usepackage{todonotes}
\usepackage{listings}
\usepackage{xcolor}
\usepackage{graphicx}

\newcommand{\code}[1]{\lstinline|#1|}

\begin{document}

\title{Chook - A comprehensive suite for generating binary optimization
problems with planted solutions}

\author{Dilina Perera}
\affiliation{Department of Physics and Astronomy, Texas A\&M University,
College Station, Texas 77843-4242, USA}

\author{Inimfon Akpabio}
\affiliation{Department of Physics and Astronomy, Texas A\&M University,
College Station, Texas 77843-4242, USA}

\author{Firas Hamze}
\affiliation{Microsoft Quantum, Microsoft, Redmond, WA 98052, USA}

\author{Salvatore Mandr\`{a}}
\affiliation{Quantum Artificial Intelligence Laboratory, NASA Ames Research Center, Moffett Field, California 94035, USA}
\affiliation{KBR, 601 Jefferson St., Houston, TX 77002, USA}

\author{Nathan Rose}
\affiliation{1QB Information Technologies Inc. (1QBit), Vancouver, BC V6C 2B5, Canada}

\author{Maliheh Aramon}
\affiliation{1QB Information Technologies Inc. (1QBit), Vancouver, BC V6C 2B5, Canada}

\author{Helmut G.~Katzgraber}
\affiliation{Microsoft Quantum, Microsoft, Redmond, WA 98052, USA}

\begin{abstract}

We present Chook, an open-source Python-based tool to generate discrete
optimization problems of tunable complexity with a priori known
solutions. Chook provides a cross-platform unified environment for
solution planting using a number of techniques, such as tile planting,
Wishart planting, equation planting, and deceptive cluster loop
planting. Chook also incorporates planted solutions for higher-order
(beyond quadratic) binary optimization problems. The support for various
planting schemes and the tunable hardness allows the user to generate
problems with a wide range of complexity on different graph topologies
ranging from hypercubic lattices to fully-connected graphs.

\end{abstract}

\maketitle

\section{Introduction}

The advent of quantum annealing devices~\cite{johnson:11-ea,bunyk:14} has
spawned a renewed interest in the development of heuristic approaches to
solve discrete optimization problems.  On the one hand, the quantum
revolution has inspired remarkable classical algorithms (see, for
example, Ref.~\cite{mandra:16b}) running on
conventional CMOS hardware that have raised the bar for emerging quantum
annealing hardware.  On the other hand, substantial progress has been
made in recent years on the development of programmable devices based on
alternative technologies such as, for example, the coherent Ising
machine based on optical parametric
oscillators~\cite{wang:13b,hamerly:18x}, digital MemComputing machines
based on self-organizing logic gates~\cite{diventra:18,traversa:15}, and
the ASIC-based Fujitsu Digital
Annealer~\cite{matsubara:17,tsukamoto:17,aramon:19}.

The rapid progression of novel computing platforms and algorithms
demands tunable hard optimization problems for benchmarking purposes.
Much effort has been devoted to generate binary synthetic benchmark
problems whose optimal configurations are known \textit{a priori}
\cite{barthel:02,hen:15a,king:15,marshall:16,hamze:18,albash:18,hen:19a,hamze:20}.
These are frequently referred to as problems with \textit{planted
solutions}.  In particular, benchmark problems that are easily scalable
to large system sizes, and ideally with tunable hardness, facilitate
systematic comparison of optimizers.  Aside from their practical
significance, theoretical investigation of such problem ensembles
reveals intriguing insights into the nature of disordered systems, in
particular, the interplay between frustration, thermodynamic behavior,
and computational complexity~\cite{hamze:20,perera:20}.

In this paper we introduce Chook (version 0.2.0), a comprehensive
Python-based tool that integrates multiple solution planting schemes to
provide a unified framework for generating benchmark problems for binary
optimization problems.  Chook currently supports tile
planting~\cite{hamze:18,perera:20}, Wishart planting~\cite{hamze:20},
deceptive cluster loops~\cite{mandra:18}, and equation
planting~\cite{hen:19a}.  In addition, the software allows for the
construction of planted solutions for problems with higher-order ($k>2$)
interactions, by combining  problems with lower-order ($k \leq 2$)
interactions.

The paper is organized as follows.  In Sec.~\ref{sec:planting_schemes}
we present an overview of the solution planting schemes included in
Chook.  Section~\ref{sec:chook} provides instructions on installation
and usage of the software, along with a detailed description of
parameters and options, followed by concluding remarks.

\section{Solution planting schemes} \label{sec:planting_schemes}

In solution planting, the goal is to construct a binary cost function
such that the minimizing configurations are known \textit{a priori}.  In
the most general form, a cost function in Ising form with variables
$\boldsymbol{s} = (s_1, \dotsc, s_N)$, $s_i \in \{\pm 1\}$ is given
by
\begin{equation} \label{eq:main_hamiltonian}
	\mathcal{H}(\boldsymbol{s}) = \sum_{j \in V} h_j s_j + \sum_{k=2}^n \; \sum_{( i_1, i_2, \dotsc, i_k ) \in E} J_{i_1 i_2 \dotso i_k} s_{i_1} s_{i_2} \dotsm s_{i_k},
\end{equation}
where the hypergraph $H = (V,E)$ with vertices $V$ and hyperedges $E$
captures the connectivity of the problem, and $\{h_j\}$ are the local
fields. A term containing a product of $k$ spins, $J_{i_1 i_2 \dotso
i_k} s_{i_1} s_{i_2} \dotsm s_{i_k}$, is referred to as a $k$-local
interaction with $J_{i_1 i_2 \dotso i_k}$ being the coupling constant.
Equation \eqref{eq:main_hamiltonian} can be readily mapped onto a cost
function of Boolean variables $\boldsymbol{x} = (x_1, x_2, \dotsc,
x_N)$, $x_i \in \{0, 1\}$ in the form of a high-order polynomial
unconstrained binary optimization (HOBO) problem via the transformation
$s_i \rightarrow 1-2x_i$.

Except for a few platforms like Azure Quantum, most of the software for 
binary optimization problems mainly targets up to 2-local interactions, 
in which case Eq.~\eqref{eq:main_hamiltonian} reduces to
\begin{equation}
	\mathcal{H}(\boldsymbol{s}) = \sum_{j \in V} h_j s_j + \sum_{(i,j) \in E} J_{ij} s_i s_j,
\end{equation}
where $G=(V,E)$ is the underlying problem graph.  Among the planting
schemes supported by Chook, tile planting, Wishart planting, and
deceptive cluster loops methods construct cost functions with $2$-local
interactions.  Equation planting and $k$-local planting methods are
capable of generating problems with higher-order ($k>2$) local
interactions.

In what follows, we provide a summary of each solution planting scheme
supported by Chook. For a detailed description, the reader is referred
to the the original references introducing the methods.

\subsection{Tile planting}

In tile planting~\cite{hamze:18} one seeks to decompose the problem
graph into edge-disjoint vertex-sharing subgraphs and embed elementary
Ising subproblems that share a common ground state over the subgraphs.
The method produces scalable problems with highly-tunable complexity on
cubic and square lattice topologies~\cite{hamze:18,perera:20}.
It also extends to arbitrary graph structures, e.g., via lattice animals.

Consider a decomposition of the problem graph $G=(V,E)$ into subgraphs
$\{ G_l =(V_l, E_l) \}$ that ensures no edges are shared among the
subgraphs (i.e., edge-disjoint). For each subgraph, we define an Ising
cost function of the form
\begin{equation}
	\mathcal{H}_l = \sum_{(i,j) \in E_l} J_{ij} s_i s_j.
\end{equation}
The subproblem Hamiltonians $\{\mathcal{H}_l\}$ are added to obtain the
complete Hamiltonian $\mathcal{H}_{_\text{TP}}$ over $G$ as
\begin{equation}
	\mathcal{H}_{_\text{TP}} = \sum_{l} \mathcal{H}_l.
\end{equation}
The subproblems are constructed such that they share a common ground
state configuration $\boldsymbol{t}$, therefore the entire problem has a
ground state characterized by the same local configuration
$\boldsymbol{t}$ occupying the constituent subgraphs. For simplicity,
$\boldsymbol{t}$ is taken to be the ferromagnetic ground state, i.e.,
$\{+1, +1, \dotsc, +1\}$ (modulo $\mathbb{Z}_2$ symmetry).  Once the
problem is constructed, the planted ferromagnetic ground state can be
concealed by a gauge randomization in which the couplers $\{J_{ij}\}$
are transformed as $J_{ij}^\prime \leftarrow J_{ij} q_i q_j$, where
$\boldsymbol{q}$ is an arbitrary ground state.

\begin{figure}[tb!] 
\includegraphics[width=0.5\linewidth]{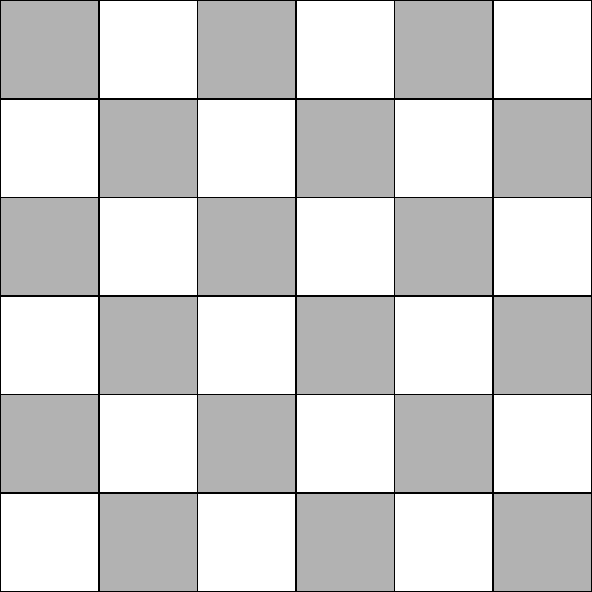}
\caption{
Decomposition of a square lattice with periodic boundary conditions into
edge-disjoint, unit-cell subgraphs (shaded cells) on which Ising
subproblems are embedded.  Under periodic boundary conditions, each
vertex is shared by exactly two subgraphs.
}
 \label{fig:checker_board}
\end{figure}

Chook supports the generation of tile-planted problems on square and
cubic lattices with periodic boundary conditions.  The regular structure
of these lattices allows for a problem-graph decomposition that
naturally renders a subset of the unit cells as the subgraphs.  Figure
\ref{fig:checker_board} illustrates this decomposition for square
lattices, in which the resulting unit-cell subgraphs (dark color) form a
checkerboard pattern.

\begin{figure}[tb!] 
\includegraphics[width=\linewidth]{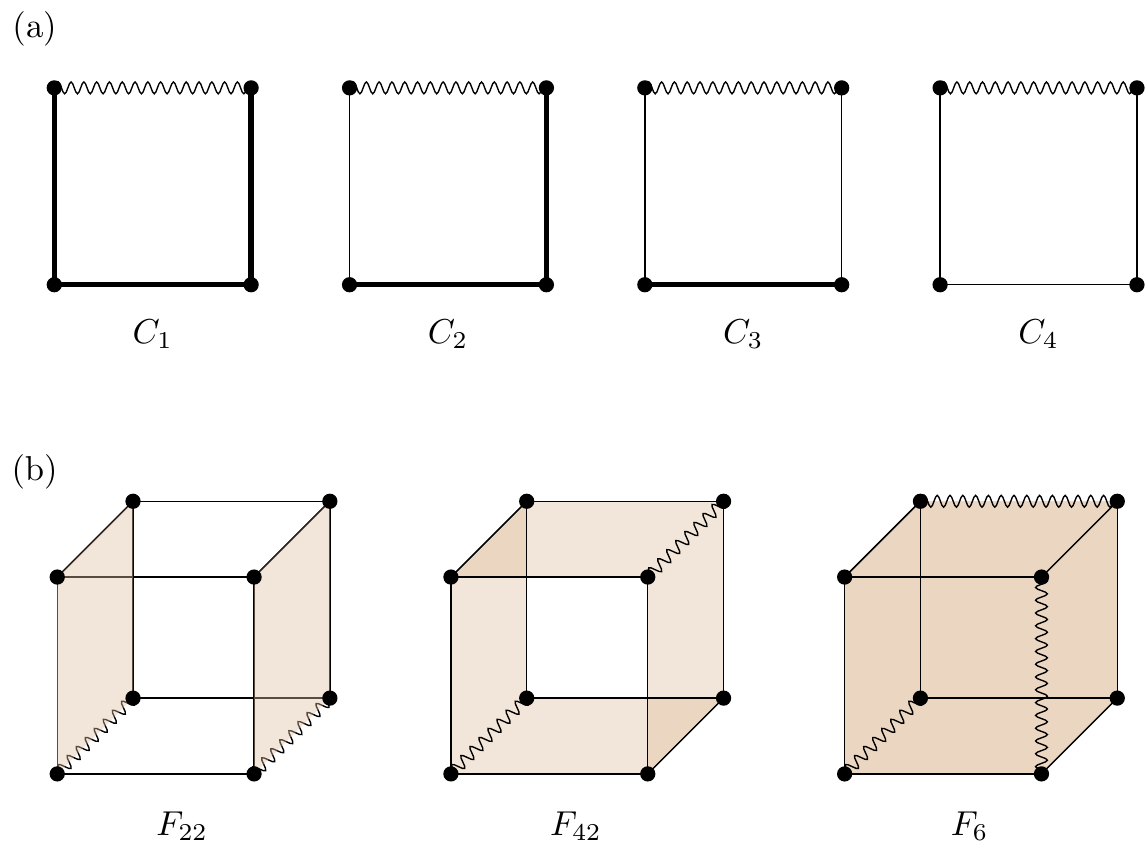}
\caption{
Class representatives of the unit-cell subproblems for (a) square-lattice
and (b) cubic-lattice topologies. Straight lines represent ferromagnetic
couplers with values $-1$ (thin lines) and $-2$ (thick lines), while
wavy lines denote antiferromagnetic couplers with value $+1$. All
subproblems have the ferromagnetic state as one of the ground states.
Classes $C_2$ and $C_3$ each have three equivalent representations (only
a single member shown) based on the ways the ferromagnetic bonds can be
distributed. $F_{22}$, $F_{42}$, and $F_6$ classes each have $48$
members that are equivalent under the octahedral symmetry.
}
\label{fig:subproblem_classes}
\end{figure}

For square lattices we define four subproblem classes $\{C_i\}$, $i \in
\{1,2,3,4\}$ that correspond to unit cycles (plaquettes) with different
levels of frustration [see Fig.~\ref{fig:subproblem_classes}(a) for an
illustration]. A subproblem from the class $C_i$ is constructed by
assigning the antiferromagnetic value $+1$ to a chosen coupler, the
ferromagnetic value $-1$ to randomly selected $i-1$ couplers, and $-2$
to the remaining couplers. Planted instances are generated by first
assigning a subproblem type for each subgraph in the lattice, followed
by a random rotation of the plaquette to allow for more disorder.  We
define instance classes based on the probability distribution over
classes $\{C_i\}$ according to which subproblem types are assigned to
subgraphs. Specifically, we denote $p_i$ to be the probability of
choosing subproblems from class $C_i$, and uniquely define each instance
class based on the three probability parameters $\{p_1, p_2, p_3\}$,
where $p_1 + p_2 + p_3 \leq 1$.  Multiple complexity transitions have
been observed in the multidimensional phase space defined by these
parameters~\cite{perera:20}.

For cubic lattices, Chook uses three subproblem types $F_{22}$,
$F_{42}$, and $F_6$, each having two, four, and six frustrated facets,
respectively [see Fig.~\ref{fig:subproblem_classes}(b)].  Each
subproblem class consists of $48$ members that are equivalent by
octahedral symmetry under the operations of rotation and
reflection~\cite{comment:subproblem_classes}. Similar to the square
lattice case, problem construction begins by assigning subproblem types
for the subgraphs according to the chosen probability distribution over
classes $\{F_{ij}\}$. Then for each subproblem type, one of the $48$
members are selected randomly. Instance classes are defined based on the
two probability parameters $\{ p_{F_{22}}, p_{F_{42}} \}$,
$p_{F_{22}}+p_{F_{42}} \leq 1$, where $p_{F_{22}}$ and $p_{F_{42}}$ are
the probabilities of choosing subproblems from the $F_{22}$ class and
$F_{42}$ class, respectively. By varying these parameters one can
achieve drastic changes in typical complexity, with higher
concentrations of $F_6$ subproblems resulting in problems that are many
orders of magnitude harder than random problems with bimodal and
Gaussian disorder~\cite{hamze:18}.

\subsection{Wishart planting}

Wishart planting~\cite{hamze:20} generates Ising Hamiltonians on
complete graphs of the form
\begin{equation}
	\mathcal{H}_{_\text{WP}}(\boldsymbol{s}) =  \frac{1}{2} \sum_{i \neq j} J_{ij} s_i s_j =  \frac{1}{2} \boldsymbol{s}^T \boldsymbol{J} \boldsymbol{s},
\end{equation}
in which the coupler matrix $\boldsymbol{J}$ follows a Wishart
distribution, a matrix generalization of the $\chi^2$ distribution.  The
model exhibits a first-order phase transition in temperature, and allows
for dramatic variations in typical hardness over many orders of
magnitude as a control parameter is varied.

In Wishart planting, one defines the coupler matrix $\boldsymbol{J}$ in
terms of a $N \times M$ real-valued matrix $\boldsymbol{W}$, where $N$
is the number of spins and $M$ ($M \ge 1$) is a tunable parameter which
regulates the typical hardness and the thermodynamic behavior of the
problem ensemble. Specifically, we define $\tilde{\boldsymbol{J}}$ as
\begin{equation}
	\tilde{\boldsymbol{J}} = \frac{1}{N} \boldsymbol{W} \boldsymbol{W}^T,
\end{equation}
and zero the diagonal to obtain $\boldsymbol{J} =
\tilde{\boldsymbol{J}}- \text{diag} (\tilde{\boldsymbol{J}})$.  It can
be shown that the problem Hamiltonian $\mathcal{H}_{_\text{WP}}$ then
becomes
\begin{equation}
	\mathcal{H}_{_\text{WP}}(\boldsymbol{s}) = \frac{1}{N} \frac{1}{2} \|\boldsymbol{W}^T \boldsymbol{s}\|^2 - \frac{1}{2} \text{Tr} (\tilde{\boldsymbol{J}}).
\end{equation} 
Because the first term is in positive semidefinite quadratic form,
$\mathcal{H}(\boldsymbol{s})$ is minimized when $\boldsymbol{s} =
\boldsymbol{t}$ at which $\boldsymbol{W}^T \boldsymbol{t} = 0$. Thus,
the goal is to construct the matrix $\boldsymbol{W}$ for a given planted
solution $\boldsymbol{t}$ such that the condition $\boldsymbol{W}^T
\boldsymbol{t} = 0$ is satisfied.  We achieve this by drawing the $M$
columns $\{\boldsymbol{\omega}^\mu\}$ of $\boldsymbol{W}$ from a
correlated Gaussian distribution ${\boldsymbol{\omega}^\mu} \sim
\mathcal{N}(0, \boldsymbol{\Sigma})$, where $\boldsymbol{\Sigma} = N[
\boldsymbol{I}_N - \boldsymbol{t} \boldsymbol{t}^T/N ]/(N-1)$ is the
covariance matrix.  More precisely, we successively sample uncorrelated
Gaussian variates ${\boldsymbol{z}^\mu} \sim \mathcal{N}(0,
\boldsymbol{I}_N)$ and then multiply by the square root of
$\boldsymbol{\Sigma}$ to obtain ${\boldsymbol{\omega}^\mu}$, i.e.,
${\boldsymbol{\omega}^\mu} = \boldsymbol{\Sigma}^{\frac{1}{2}}
\boldsymbol{z}^\mu$.  It can be shown that $\langle
\boldsymbol{\omega}^\mu, \boldsymbol{t} \rangle = 0$ for all $\mu$
generated in this manner. The matrix $\boldsymbol{W} \boldsymbol{W}^T$
is known to follow a Wishart distribution. Analogous to the
clause-to-variable ratio in Boolean satisfiability problems, we define
an equation-to-variable ratio $\alpha = M/N$ for modulating the
typical computational hardness of the problem ensemble. The class
exhibits a pronounced easy-hard-easy complexity transition as $\alpha$
is varied~\cite{hamze:20}.

One can construct problems with discrete coupler values via a simple
modification of the sampling procedure. When generating
$\{\boldsymbol{\omega}^\mu\}$, instead of using Gaussian variates
${\boldsymbol{z}^\mu} \sim \mathcal{N}(0, \boldsymbol{I}_N)$, one can
sample from a Rademacher distribution, i.e., draw samples independently
and uniformly from $\{-1, 1\}$. It can be shown that the scaled couplers
$J_{ij}^\prime = N^2 (N-1) J_{ij}$ assume values in the set of
equally-spaced integers given by
\begin{equation}
J_{ij}^\prime \in \{ -4M(N-1)^2, \dotsc, -4, 0, 4, \dots, 4M(N-1)^2 \}.
\end{equation}

\subsection{Deceptive cluster loops}

Deceptive cluster loops (DCL)~\cite{mandra:18} are a class of benchmark
problems designed for the Chimera topology of the D-Wave 2000Q and
D-Wave 2X quantum annealers, although the ideas can be generalized to
other topologies. DCL problems are derived from the conventional
frustrated cluster loops (FCL) problems~\cite{king:17}.
They have an additional control parameter that conceals the logical 
structure of the problem for a particular range of values. Such a 
feature makes these problems hard to solve using algorithms that exploit 
the underlying logical structure.

Both DCL and FCL problems are based on the traditional frustrated loop
problems introduced in Ref.~\cite{hen:15a}. Frustrated loop problems are
constructed by generating $M = \alpha N$ loops on a graph $G=(V,E)$,
where $N$ is the number of nodes and $\alpha$ is the loop-to-node ratio.
The loops are generated by placing random walkers on random nodes. A
random walk is terminated when it crosses its own path, and the trailing
tail segment is discarded to form a closed loop. On each loop $k$, the
coupler values $J_{ij}^k$ are set to the ferromagnetic value $-1$ except
for a single, randomly-chosen coupler for which we assign the
antiferromagnetic value $+1$. The total Hamiltonian is formed by adding
up the couplers belonging to all the loops
\begin{equation}
	\mathcal{H}_{_\text{DCL}} = \sum_{(i,j) \in E} \; \sum_{k=1}^M J_{ij}^k s_i s_j.
\end{equation}
The problem is discarded if $|\sum_{k=1}^M J_{ij}^k| > R$ for any edge
$(i,j) \in E$, where $R$ is referred to as the ``ruggedness.''  In
frustrated cluster loop (FCL) problems~\cite{king:17}, frustrated loops
are generated on a two-dimensional logical lattice embedded on a Chimera
graph. Here, all couplers within each Chimera unit cell are set to the
ferromagnetic value $-1$, forcing all physical spins within the unit
cell to behave as a logical spin. The frustrated loops are then
generated on the resulting $L_x \times L_y$ two-dimensional logical
lattice, where $L_x$ and $L_y$ are the linear dimensions of the parent
Chimera graph.

DCL problems are an extension of FCL problems. Here, all inter-cell
couplers between Chimera unit cells are multiplied by a scaling factor
$\lambda$, while leaving all intra-cell couplers intact.  For small
values of $\lambda$ ($\lambda \sim 1$), DCL problems can be described by
a virtual planer model with each Chimera unit-cell behaving as a single
virtual variable. For $\lambda \gg 1$, the model behaves as a virtual
fully-connected bipartite problem.  For intermediate values of
$\lambda$, the problem behavior is nontrivial and no logical structures 
(as it happens for either small or large values of $\lambda$) can be 
determined to simplify the optimization of the problem; see 
Ref.~\cite{mandra:18} for details.

\subsection{Equation planting}

Systems of linear equations modulo 2, also known as ``exclusive or
satisfiability'' (XORSAT) equations, are solvable in polynomial time by
Gaussian elimination, but when formulated as optimization problems, they
are often challenging for heuristic
solvers~\cite{joerg:10,guidetti:11,farhi:12}.  Equation
planting~\cite{hen:19a} casts XORSAT problems as $k$-local ($k>1$) Ising
cost functions.

Consider a linear system of equations modulo 2 with $N$ Boolean
variables $\{x_1, x_2, \dots, x_N\}$ and $M$ equations
\begin{equation}
\sum_{j=1}^N a_{ij} x_j = b_i, 
\end{equation}
for $i \in \{1, \dotsc, M \}$, where the coefficients $\{a_{ij},
b_i\} \in \{0, 1\}$.  Alternatively, each equation can be written in
terms of the bit-wise XOR operation as
$a_{i1} x_1 \oplus \dotsm \oplus a_{iN} x_N = b_i$.
When expressed in the Ising form, the above equation becomes
\begin{equation}
\prod_{j: a_{ij}=1} s_j = (-1)^{b_i} \qquad s_j \in \{\pm 1\},
\end{equation}  
in which only the variables with non-zero coefficients are included in
the product.  The linear system can be cast as an optimization problem
with an Ising cost function, i.e., 
\begin{equation}
\mathcal{H}^\prime = \sum_{i=1}^M \left[  \prod_{j: a_{ij}=1} s_j - (-1)^{b_i}  \right]^2,
\end{equation}
which, after dropping irrelevant constants, reduces to
\begin{equation}
\mathcal{H}_{_\text{XORSAT}} = -\sum_{i=1}^M (-1)^{b_i}  \prod_{j: a_{ij}=1} s_j.
\end{equation} 
The ground state of the Ising cost function corresponds to the solution
of the linear system, given that the linear system is solvable.  Here we
limit our attention to $k$-regular $k$-XORSAT problems, where each
equation contains exactly $k$ randomly selected variables out of the $N$
variables (hence $k$-XORSAT), and each variable appears in exactly $k$
equations (hence $k$-regular).  Such a linear system consists of $N$
equations (i.e., $M=N$), and the resultant Ising Hamiltonian contains
$N$ $k$-local terms. If the linear system has solutions, the
ground-state energy of the Ising cost function is $-M$, and the
ground-state degeneracy (i.e., the number of minimizing configurations)
is given by $g = 2^{N-r}$, where $r$ is the number of linearly independent
(in arithmetic modulo 2) rows of the matrix $\boldsymbol{A}=(a_{ij})$. 
For random $k$-regular $k$-XORSAT instances, $r$ is typically $\mathcal{O}(N)$
and hence the number of ground states grows sub-exponentially
~\cite{mandra:16c, alamino:09, alamino:08}.

\subsection{$k$-local planting}

Here we introduce a method of planting solutions for Hamiltonians with
higher-order ($k>2$) interactions by combining Hamiltonians with
lower-order ($k \leq 2$) interactions and known ground states.

Consider a set of $n$ problems described by the Hamiltonians $\{
\mathcal{H}^{(1)}(\boldsymbol{s}^{(1)}),
\mathcal{H}^{(2)}(\boldsymbol{s}^{(2)}), \dots,
\mathcal{H}^{(n)}(\boldsymbol{s}^{(n)}) \}$ whose ground state energies
are $\{E_0^{(1)}, E_0^{(2)}, \dots, E_0^{(n)}\}$.   
The composite Ising cost function defined by the product
\begin{equation}
\mathcal{H}_{_\text{comp}} = \prod_{i=1}^n \left[ 
\mathcal{H}^{(i)} - E_0^{(i)} \right]
\end{equation}
is minimized for $\boldsymbol{s}^{(i)} = \boldsymbol{t}^{(i)}$, $i
\in \{1, \dotsc, n\}$, where $\boldsymbol{t}^{(i)}$ is a ground state of
$\mathcal{H}^{(i)}$.  
To reduce the total number of variables in the composite problem,
we allow spins to be shared among subproblems.
The composite problem hence contains 
$N = \text{max} \, \{ N^{(1)}, N^{(2)}, \cdots, N^{(n)}\}$ spins, 
where $N^{(i)}$ is the number of spins in the $i$th problem. 
The locality of the highest-order term in the composite
Hamiltonian $\mathcal{H}_{_\text{comp}}$ is $k_\text{max} = \sum_{i=1}^n
k_\text{max}^{(i)}$, with $k_\text{max}^{(i)}$ being the locality of the
highest order term in $\mathcal{H}^{(i)}$.

Chook constructs higher-order ($k>2$) cost functions with even
$k_\text{max}$ by using tile-planted problems and Wishart-planted
problems as constituent problems \cite{comment:precision}. 
Because these problem types consist of
only $2$-local interactions, one cannot construct composite problems
with odd $k_\text{max}$ using these two problem types alone. Therefore
when generating problems with odd $k_\text{max}$, we also use a trivial
Hamiltonian with $1$-local interactions, namely, Ising spins coupled to
a bimodal random field, given by $\mathcal{H}_\text{1-local} =
\sum_{i=1}^{N_l} h_i s_i$, where $N_l$ is the number of spins and
$\{h_i\}$ are the random fields drawn independently and uniformly from
$\{+1, -1\}$.  The ground state of $\mathcal{H}_\text{1-local}$ is
trivially obtained by aligning all spins with their random fields.

\section{Chook} 
\label{sec:chook}

Chook is a platform-independent, Python-based tool distributed under the
Apache License 2.0.  The code is openly available on the GitHub software
sharing platform (\code{https://github.com/dilinanp/chook}), as well as 
being included in this submission (see ancillary files). It is also hosted 
on the Python Package Index (PyPI) for easy installation via the package 
management system pip. We ask you to please cite this work if you choose 
to use Chook.

\subsection{Installation}

Chook requires Python version 3.4 or above for installation.  We
recommend installing Chook in a Python virtual environment to avoid
potential conflicts with packages installed system-wide.  If the Python
package management system pip is available, Chook can be installed from
the command line by running the command:
\begin{lstlisting}
	$ pip install chook
\end{lstlisting}
Alternatively, chook can be directly downloaded from PyPI or the GitHub
repository and can be installed with the command:
\begin{lstlisting}
	$ python setup.py install 
\end{lstlisting}
During installation, the prerequisite Python libraries numpy, scipy, and
more-itertools will be automatically installed if they are not already
available.

\subsection{Usage}

Chook can be executed from the command line providing two required
positional arguments \code{<problem_type>} and \code{<config_file>},
followed by zero or more optional arguments.  The basic usage ignoring
the optional arguments is
\begin{lstlisting}
	$ chook <problem_type> <config_file>
\end{lstlisting}
where the different types of \code{<problem_type>} 
are listed in Table~\ref{tab:problem_types},
and \code{<config_file>} is a configuration file in INI format that
contains the problem-type specific parameters (see
Sec.~\ref{sec:config_file}). Table~\ref{tab:arguments} shows the
complete list of optional arguments that can be appended to change the
default behavior of Chook.

\begin{table}
\caption{
Solution planting types supported by Chook and the
positional argument \code{<problem_type>} which is also the section
header in the configuration file under which the problem-type specific
parameters are listed. Note that for the problem type, the brackets
should be removed, e.g., for a tile-planted problem the \code{<problem_type>} 
is \code{TP} and the section header \code{[TP]}.
\label{tab:problem_types}}
\begin{tabular*}{\columnwidth}{@{\extracolsep{\fill}}  l l }
\hline \hline 
Problem type & INI section header  \\ [0.5ex] 
\hline
Tile planting & \code{[TP]} \\
Wishart planting & \code{[WP]} \\
Deceptive cluster loops & \code{[DCL]} \\
Equation planting & \code{[XORSAT]} \\
$k$-local planting & \code{[K\_LOCAL]} \\
\hline
\hline
\end{tabular*}
\end{table}

\begin{table*}
\centering
\caption{Optional command-line arguments supported by Chook.
	\label{tab:arguments}}
\begin{tabular}{@{\extracolsep{\fill}} p{4cm} p{12cm}}
\hline \hline 
Argument flag & Description \\ [0.5ex] 
\hline
\code{-n <num_instances>} & \code{<num_instances>} is the number of instances to be generated (default value: 10). \\
\code{-o <output_format>} & \code{<output_format>} can either be \code{ising} or \code{hobo}, 
                            and specifies whether the problem is expressed in
the Ising form or the binary HOBO form (default value: \code{ising}).\\
\code{-f <file_format>}   & \code{<file_format>} can either be \code{txt} or \code{json}, 
                            and specifies whether the output files are in text
format or the JSON format (default value: \code{txt}). \\
\code{-h, --hel}\code{p}  & Show a help message and exit.\\
\hline
\hline

\end{tabular}
\end{table*}

\subsection{Output}

Chook stores the generated problem instances in a subdirectory under the
current working directly, which is named according to the problem type
and major problem-specific parameters. For example, tile-planted problems on
a square lattice with linear lattice size $L=32$ and tuning parameters
$p_1=0.2$, $p_2=0.5$, and $p_3=0.1$ will be stored in a subdirectory
named \code{tile_planting_2D_L_32_p1_0.2_p2_0.5_p3_0.1}. Each problem
instance will be stored in a separate file within the said subdirectory.
Except for deceptive cluster loop problems for which the ground states
cannot be decoded (except in specific limits), the ground-state energies
will be stored in a separate file \code{gs_energies.txt} with two
columns: the instance file name and the ground state energy. For the
case of XORSAT problems, \code{gs_energies.txt} will contain a third
column, the ground state degeneracy.

The generated instances are expressed in the Ising form by default,
but can be cast in the binary HOBO form by setting the optional argument
\code{-o hobo}. By default, the instance files are in plain text format.
Each line in the instance file corresponds to a single $k$-local
interaction term following the format
\begin{lstlisting}
	<i_1>  <i_2>  ...  <i_k>  <J>
\end{lstlisting}
where the first $k$ elements are the indices of the spin (Boolean)
variables, and the last element \code{<J>} is the coupling constant.
For the case of field terms (i.e., $1$-local interactions), the
entry consists of two terms: the index of the spin (Boolean) variable
followed by the field value. The user has the option to save the
instances in the JavaScript Object Notation (JSON) format by setting the
optional argument \code{-f json}.

\subsection{Configuration file} 
\label{sec:config_file}

As the second command-line argument, Chook requires a configuration file
in INI format that specifies the parameters associated with the selected
problem type.  Chook is distributed with a sample configuration file
\code{params.i}\code{n} which includes sample parameter specifications
for all problem types. Each parameter is specified as an INI property
with a \code{name} and a \code{value}, separated by an equal sign, i.e.,
\code{name = value}. Properties are grouped into sections according to
the problem types they are associated with, and the sections begin with
section headers of the form \code{[<problem_type>]}.
Table~\ref{tab:problem_types} shows the section headers associated with
the supported problem types. A section ends when the next section
header is encountered, or when the end of file is reached. When Chook
is executed, only the section associated with the chosen problem type is
read from the configuration file, and the rest of the file is ignored.
Table~\ref{tab:properties} shows a comprehensive list of supported
problem-specific properties.  We recommend the users to modify the
provided \code{params.i}\code{n} file to meet their needs rather than
scripting their own configuration file to avoid potential errors.

We now describe the parameter specification for the $k$-local planting
method.  In Chook, $k$-local problems are constructed by combining a
sequence of constituent problems (or ``subproblems'') of three supported
types, namely, tile planting, Wishart planting, and bimodal
random field terms.  Parameter specifications for the subproblems should
be grouped into separate INI sections with user-defined headers and
should be listed below the main section \code{[K_LOCAL]}.  The section
\code{[K_LOCAL]} contains two properties: \code{k_max} which represents
the locality $k_\text{max}$ of the highest-order term in the composite
Hamiltonian, and \code{subproblem_id_list} that accepts a list of
identifiers representing the subproblems.  Each subproblem identifier
\code{<subproblem_id>}, enclosed within square brackets, i.e.,
\code{[<subproblem_id>]}, is used as the header of the section under
which the subproblem properties are listed.  Subproblem identifiers
should begin with a letter, and may contain a combination of letters and
numbers.  One can reuse the same subproblem (with the same parameter
specifications) multiple times in the construction procedure, in which
case the corresponding subproblem identifier should be repeated
accordingly in the \code{subproblem_id_list}.  In addition to the usual
problem-specific properties, each section defining a subproblem should
contain an additional property \code{subproblem_type}, which is used to
identify the type of the subproblem being defined.  The allowed values
are, ``\code{RF}'' for the bimodal random field terms, ``\code{TP}'' for tile planting,
and ``\code{WP}'' for Wishart planting.  As the bimodal random field terms can
be trivially optimized, Chook enforces the user to minimize its usage,
and it is only allowed when constructing problems with odd values of
$k_\text{max}$.  Therefore, for even values of $k_\text{max}$, one should use
$k_\text{max}/2$ subproblems with $2$-local interactions, which can be
tile-planted problems and/or Wishart problems. For odd values of
$k_\text{max}$, one should use $(k_\text{max}-1)/2$ subproblems with
$2$-local interactions, and a single subproblem with a $1$-local term
(i.e., a bimodal random field term).

\begin{table*}[t]
\centering
\caption{Problem-specific properties defined in the configuration file.
         Note that the term ``variables'' refers to either spins (in Ising form) or Boolean variables (in HOBO form).
	\label{tab:properties}}
\begin{tabular}{@{\extracolsep{\fill}} p{3cm} p{4cm} p{10.5cm}}
\hline \hline 
Section header & Property name & Description \\ [0.5ex] 
\hline
\multirow{14}{4em}{\code{[TP]}} 
& \code{dimension}       & Spatial dimension $D$ of the periodic lattice on which problems are constructed. 
                           Should be set to $2$ for a square-lattice and $3$ for a cubic-lattice. \\
& \code{L}               & Linear lattice size $L$. 
                           Must be an even integer greater than two. 
                           The number of variables in the planted problem is $N=L^D$.\\ 
& \code{p1, p2, p3}      & Probabilities $\{p_1, p_2, p_3\}$ for constructing problems on square-lattice topology. 
                           $p_i$ is the probability of choosing subproblems from class $C_i$.
                           Should satisfy the conditions $p_1+p_2+p_3 \leq 1.0$ and $0 \leq p_i \leq 1$, $i \in \{1,2,3\}$. 
					      These parameters are ignored when \code{dimension = 3}. \\
& \code{	pF22, pF42}      & Probabilities $\{p_{F_{22}}, p_{F_{42}}\}$ for constructing problems on a cubic-lattice.
                           $p_{F_{22}}$ and $p_{F_{42}}$ are the probabilities of selecting subproblems from classes $F_{22}$ and
                           $F_{42}$, respectively.
    					      Should satisfy the conditions $p_{F_{22}} + p_{F_{42}} \leq 1.0$ 
    					      and $0 \leq p_{F_{22}}, p_{F_{42}} \leq 1$. 
    					      These parameters are ignored when \code{dimension = 2}. \\		
& \code{gauge_transform} & If set to \code{yes}, the planted ferromagnetic ground state will be concealed
						  via a gauge randomization. Allowed values: \{\code{yes}, \code{no}\} \\ 	
\hline
\multirow{10}{4em}{\code{[WP]}} 
& \code{N}                   & The number of variables $N$ in the planted problem. \\
& \code{alpha}               & Equation-to-variable ratio $\alpha=M/N$, where $M$ is the
                               number of columns in the matrix $\boldsymbol{W}$.
                               Note that $M$ is determined from $\alpha$ as $M=\alpha N$, and will be rounded to the nearest 
                               non-zero integer.
                               Thus the value of $\alpha$ internally represented by Chook can be different
                               from the user-provided value. \\
& \code{discretize_couplers} & If set to \code{yes}, the code generates problems with discrete couplers by
							  sampling from a Rademacher distribution instead of a Gaussian distribution
							  when constructing the matrix $\boldsymbol{W}$.
							  Allowed values: \{\code{yes}, \code{no}\} \\
& \code{gauge_transform}     & If set to \code{yes}, the planted ferromagnetic ground state will be concealed
						      via a gauge randomization. Allowed values: \{\code{yes}, \code{no}\} \\  
\hline
\multirow{13}{4em}{\code{[DCL]}}
& \code{Lx, Ly}   & Linear dimensions $L_x$ and $L_y$ of the Chimera graph 
                    on which problems are constructed ($ 1 \leq L_x, L_y \leq 16$) .
                    Frustrated loops are generated on the $L_x \times L_y$ logical lattice which treats
                    Chimera unit cells as logical variables.
                    The number of (physical) variables is $N = 8 L_x L_y$. \\ 
& \code{alpha}    & Loop-to-node ratio $\alpha$ defined by $\alpha=M/N_l$, where $N_l = L_x \times L_y$ 
                    is the number of logical variables,
                    and $M$ is the number of loops generated on the logical lattice.
                    Note that $M$ is determined from $\alpha$ as $M=\alpha N_l$, and will be rounded to the nearest 
                    non-zero integer.
                    Thus the value of $\alpha$ internally represented by Chook can be different
                    from the user-provided value. \\
& \code{R}        & Ruggedness $R$ that limits the range of coupler strength as $|\sum_{k=1}^M J_{ij}^k| > R$.
                    Must be an integer greater than zero.\\
& \code{lambda}   & Scaling factor $\lambda \geq 1$ by which the inter-cell couplers between Chimera
                    unit cells are scaled.
                    For $\lambda = 1$, DCL problems are equivalent to FCL problems. \\

\hline
\multirow{4}{4em}{\code{[XORSAT]}}
& \code{k} & Locality $k$ of the terms in the resultant Ising cost function 
			(equivalent to the number of variables per equation in $k$-regular $k$-XORSAT).\\
& \code{N} & The total number of variables $N$ in the problem ($N \geq k$). 
			This is equivalent to the number of equations in the $k$-regular $k$-XORSAT representation. \\
			
\hline
\multirow{8}{4em}{\code{[K_LOCAL]}}
& \code{k_max}              & Locality $k_\text{max}>2$ of the highest-order term 
                              in the composite Hamiltonian. \\
& \code{subproblem_id_list} & A list of user-defined, comma-delimited identifiers representing the subproblems.
                              Subproblem identifiers should begin with a letter and can include alphanumeric characters.
                              Each identifier enclosed by square brackets, i.e., \code{[<subproblem_id>]}, 
                              is used as the section header under which the properties of the subproblem are grouped.
                              A subproblem identifier can be repeated multiple times in \code{subproblem_id_list},
                              if one chooses to reuse the corresponding subproblem specification multiple times.\\
                              
\hline
\multirow{4}{4em}{\code{[<subproblem_id>]}}
& \code{subproblem_type}                  & A numeric code used to identify the type of subproblem being defined. 
                                            Should be included with every INI section defining a subproblem,
                                            in addition to the subproblem-specific properties. 
                                            Allowed values: 
                                            \code{RF} -- Bimodal random field , \code{TP} -- Tile planting, \code{WP} -- Wishart planting \\
& \code{<subproblem-specific properties>} & Define all the properties associated with the selected subproblem type.  
                                            If the subproblem type is a bimodal random field, define a single property
                                            \code{N}, which specifies the number of variables in the subproblem. \\

\hline
\hline

\end{tabular}
\end{table*}

\section{Summary}
\label{sec:summary}

We have presented a Python-based suite, Chook, for generating discrete
optimization problems with known solutions using multiple popular
solution planting schemes.  Chook is distributed freely via Python
Package Index (PyPI) and GitHub software sharing platform, and allows
for fast and easy installation on any platform.  The code supports the
construction of cost functions with tunable hardness and local
interactions spanning $2$-local to arbitrary higher-order.  We believe
that Chook will be highly beneficial for generating benchmark problem
sets for current and future generations of programmable devices for
discrete optimization.

\begin{acknowledgments} 

D.P. would like to acknowledge Chris Pattison for suggestions during the
initial code development phase.  H.G.K.~would like to thank Nacho Kilber
for inspiration.  F.H. is indebted to Julia Child for stimulating his
work in this area through her classic and pioneering oeno-thermalized
Chook algorithm.  Part of this research is based upon work supported in
part by the Office of the Director of National Intelligence (ODNI),
Intelligence Advanced Research Projects Activity (IARPA), via MIT
Lincoln Laboratory Air Force Contract No.~FA8721-05-C-0002.  The views
and conclusions contained herein are those of the authors and should not
be interpreted as necessarily representing the official policies or
endorsements, either expressed or implied, of ODNI, IARPA, or the
U.S.~Government.  The U.S.~Government is authorized to reproduce and
distribute reprints for Governmental purpose notwithstanding any
copyright annotation thereon.

\end{acknowledgments}

\bibliography{refs,comments}

\begin{thebibliography}{29}
\bibitem[{com({\natexlab{a}})}]{comment:subproblem_classes}
\bibinfo{note}{One can define two additional subproblem classes, $F_{21}$ and
  $F_{41}$, having the same number of frustrated facets as the $F_{22}$ and
  $F_{42}$ classes, respectively. Each of these classes also consists of 48
  members that are equivalent under octahedral symmetry.}
\bibitem[{com({\natexlab{b}})}]{comment:precision}
\bibinfo{note}{Note that when multiplying Wishart problems to generate
  higher-order Hamiltonians the range of the couplers grows, thus making the
  problems considerably harder to solve. We recommend using sub-problem classes
  with integer coupler values to avoid this problem.}
\end{thebibliography}

\end{document}